\documentclass[twocolumn,twocolappendix]{aastex631}
\graphicspath{{./fig/}{./png/}}
\usepackage{bm,color}

\newcommand{\EQ}{\begin{equation}}
\newcommand{\EN}{\end{equation}}
\newcommand{\EQA}{\begin{eqnarray}}
\newcommand{\ENA}{\end{eqnarray}}

\newcommand{\Eq}[1]{Equation~(\ref{#1})}
\newcommand{\Eqs}[2]{Equations~(\ref{#1}) and~(\ref{#2})}

\newcommand{\Sec}[1]{Sect.~\ref{#1}}

\newcommand{\Fig}[1]{Figure~\ref{#1}}

\newcommand{\Tab}[1]{Table~\ref{#1}}

\newcommand{\bra}[1]{\langle #1\rangle}
\newcommand{\bbra}[1]{\left\langle #1\right\rangle}

{}
{}
{}

{}
{}
{}
{}
{}
{}
{}
{}
{}
{}
{}
{}
{}
{}
{}
{}
{}

{}

\newcommand{\hatkk}{\hat{\bm{k}}}

{}

{}
{}
{}

\newcommand{\ww}{\mbox{\boldmath $w$} {}}

\newcommand{\kk}{\bm{k}}

\newcommand{\xx}{\bm{x}}

\newcommand{\BB}{\bm{B}}

\newcommand{\JJ}{\bm{J}}

\newcommand{\AAA}{\bm{A}}

\newcommand{\FF}{\bm{F}}

\newcommand{\uu}{\bm{u}}

\newcommand{\nab}{{\bm{\nabla}}}
\newcommand{\GG}{\mbox{\boldmath $G$} {}}

\newcommand{\SSSS}{\mbox{\boldmath ${\sf S}$} {}}

\newcommand{\ii}{{\rm i}}

\newcommand{\DD}{{\rm D} {}}

\newcommand{\dd}{{\rm d} {}}

\newcommand{\const}{{\rm const}  {}}

\def\la{\mathrel{\mathchoice {\vcenter{\offinterlineskip\halign{\hfil
$\displaystyle##$\hfil\cr<\cr\sim\cr}}}
{\vcenter{\offinterlineskip\halign{\hfil$\textstyle##$\hfil\cr<\cr\sim\cr}}}
{\vcenter{\offinterlineskip\halign{\hfil$\scriptstyle##$\hfil\cr<\cr\sim\cr}}}
{\vcenter{\offinterlineskip\halign{\hfil$\scriptscriptstyle##$\hfil\cr<\cr\sim\cr}}}}}
\def\ga{\mathrel{\mathchoice {\vcenter{\offinterlineskip\halign{\hfil
$\displaystyle##$\hfil\cr>\cr\sim\cr}}}
{\vcenter{\offinterlineskip\halign{\hfil$\textstyle##$\hfil\cr>\cr\sim\cr}}}
{\vcenter{\offinterlineskip\halign{\hfil$\scriptstyle##$\hfil\cr>\cr\sim\cr}}}
{\vcenter{\offinterlineskip\halign{\hfil$\scriptscriptstyle##$\hfil\cr>\cr\sim\cr}}}}}
\def\H{\mbox{\rm H}}

\def\Ma{\mbox{\rm Ma}}

\def\Pm{\mbox{\rm Pr}_{\rm M}}

\def\Rey{\mbox{\rm Re}}

\def\EEA{{\cal E}_{\rm A}}
\def\EEV{{\cal E}_{\rm V}}
\def\EEK{{\cal E}_{\rm K}}

\def\EEM{{\cal E}_{\rm M}}

\def\EA{E_{\rm A}}
\def\EV{E_{\rm V}}
\def\EK{E_{\rm K}}
\def\EM{E_{\rm M}}

\def\cs{c_{\rm s}}
\def\CA{C_{\rm A}}
\def\CV{C_{\rm V}}
\def\CK{C_{\rm K}}
\def\CM{C_{\rm M}}

\def\xiK{\xi_{\rm K}}

\def\vA{v_{\rm A}}

\def\vAz{v_{\rm A0}}
\def\vAo{v_{\rm A1}}

\def\EM{E_{\rm M}}

\def\epsA{\epsilon_{\rm A}}
\def\epsV{\epsilon_{\rm V}}
\def\epsK{\epsilon_{\rm K}}
\def\epsM{\epsilon_{\rm M}}

\def\Brms{B_{\rm rms}}

\def\urms{u_{\rm rms}}

\def\half{{\textstyle{1\over2}}}

\def\onethird{{\textstyle{1\over3}}}

\def\fourthird{{\textstyle{4\over3}}}

\newcommand{\W}{\,{\rm W}}
\newcommand{\V}{\,{\rm V}}

\newcommand{\T}{\,{\rm T}}
\newcommand{\G}{\,{\rm G}}

\newcommand{\K}{\,{\rm K}}
\newcommand{\g}{\,{\rm g}}
\newcommand{\s}{\,{\rm s}}

\newcommand{\m}{\,{\rm m}}

\newcommand{\J}{\,{\rm J}}

\newcommand{\A}{\,{\rm A}}

\def\apjl{ApJL}
\def\apjs{ApJS}

\def\aap{A\&A}

\received{2025, January 30}
\revised{2025, March 4}
\accepted{2025, March 5}

\begin{document}

\title{Magnetically Assisted Vorticity Production in Decaying Acoustic Turbulence}

\author[0000-0002-7304-021X]{Axel Brandenburg}
\affiliation{Nordita, KTH Royal Institute of Technology and Stockholm University, Hannes Alfv\'ens v\"ag 12, SE-10691 Stockholm, Sweden}
\affiliation{The Oskar Klein Centre, Department of Astronomy, Stockholm University, AlbaNova, SE-10691 Stockholm, Sweden}
\affiliation{McWilliams Center for Cosmology \& Department of Physics, Carnegie Mellon University, Pittsburgh, PA 15213, USA}
\affiliation{School of Natural Sciences and Medicine, Ilia State University, 3-5 Cholokashvili Avenue, 0194 Tbilisi, Georgia}

\author[0000-0002-3193-1196]{Evan Scannapieco}
\affiliation{School of Earth and Space Exploration, Arizona State University, P.O.\ Box 876004, Tempe, AZ 85287, USA}

\begin{abstract}

We study vorticity production in isothermal, subsonic, acoustic (nonvortical), and decaying turbulence due to the presence of magnetic fields. Using three-dimensional numerical simulations, we find that the resulting kinetic energy cascade follows the ordinary Kolmogorov phenomenology involving a constant spectral energy flux. The nondimensional prefactor for acoustic turbulence is larger than the standard Kolmogorov constant due to the inefficient dissipation of kinetic energy. We also find that the Lorentz force can drive vortical motions even when the initial field is uniform, by converting a fraction of the acoustic energy into vortical energy. This conversion is shown to be quadratic in the magnetic field strength and linear in the acoustic flow speed. By contrast, the direct production of vortical motions by a non-force-free magnetic field is linear in the field strength. Our results suggest that magnetic fields play a crucial role in vorticity production in cosmological flows, particularly in scenarios where significant acoustic turbulence is prevalent. We also discuss the implications of our findings for the early Universe, where magnetic fields may convert acoustic turbulence generated during cosmological phase transitions into vortical turbulence.
\end{abstract}


\section{Introduction} \label{sec:intro}

One can envisage diverse astrophysical situations where the velocity field is irrotational and the gas motions are predominantly acoustic.  Such flows can be described as the gradient of a potential function, and thus may arise from  gravitational accelerations or barotropic pressure fluctuations.

Vortical motions, on the other hand, arise hydrodynamically through shocks \citep{Porter+15} and through the baroclinic term resulting from oblique gradients of density and pressure \citep{DSB11, Federrath+11, Jahanbakhshi+15, Elias-Lopez+23, Elias-Lopez+24}. However, the efficiency of these effects is limited because they depend on the Mach number, which is often small. Thus, thermal effects such as differential heating may be too weak to produce baroclinicity.

On the other hand, it has been known for some time that magnetic fields create vorticity regardless of the possible presence of irrotational turbulence as long as the curl of the Lorentz force is nonvanishing. This was demonstrated by \cite{Kahn+12}, who were primarily interested in the effect of turbulence from cosmological phase transitions on an inflationary-generated magnetic field. Yet the possibility of producing vorticity in the presence of magnetic fields is more general and may also have occurred under other circumstances.

A characteristic property of vortical turbulence is the constancy of the energy flux from the driving scale along the turbulent cascade down to the dissipation scale. This allows one to express the energy spectrum in nondimensional form, yielding a dimensionless prefactor known as the Kolmogorov constant \citep{Frisch95,Sreenivasan95}, which is well measured in vortical turbulence.

There have been numerous studies of acoustic turbulence, starting with the early works of \cite{KP73}, \cite{Elsasser+Schamel74,Elsasser+Schamel76}, and \cite{Lvov+Mikhailov81}. However, many subsequent studies focused on compressibility effects \citep{Passot+Pouquet87, Shivamoggi92, Cho+Lazarian05, Galtier+Banerjee11}.
Although the spectral properties of acoustic turbulence have also been investigated in some detail \citep{Falkovich+Meyer96, Kuznetsov+Krasnoselskikh08, Kochurin+Kuznetsov22, Ricard+Falcon23}, no values for a Kolmogorov-like prefactor have been quoted for magnetized acoustic turbulence.

Here, we use numerical simulations to revisit magnetic vorticity production in acoustic turbulence, focusing on three main questions:
(1) can the Kolmogorov prefactor be determined for acoustic turbulence, and how does the presence of a magnetic field change its value?
(2) To what extent does magnetically modified acoustic turbulence resemble ordinary turbulence?
(3) Can acoustic turbulence be converted into vortical turbulence due to the presence of a magnetic field?

We emphasize that we are not concerned with strong compressibility effects, which would occur at large Mach numbers \citep{Schleicher+13,Federrath+14, Porter+15}. This is why we prefer the term ``acoustic'' \citep{KP73} over ``compressive.'' Furthermore, compared to the more general term ``irrotational,'' the term ``acoustic'' is more directly suggestive of low-amplitude subsonic flows.

The structure of this work is as follows. In $\S2$ we describe the magnetohydrodynamic (MHD) equations and their implications for vorticity production, as well as our numerical approach, parameter space, and analysis techniques. In $\S3$ we present our results, focusing on measurements of the Kolmogorov prefactor and magnetic vorticity production. Implications and conclusions are given in $\S4.$

\section{The Model} \label{sec:model}

\subsection{Basic Equations} \label{sec:equations}

We solve the hydrodynamic and MHD equations with an isothermal equation of state, where the pressure $p$ and density $\rho$ are related to each other through $p=\rho\cs^2$ with $\cs=\const$ being the isothermal speed of sound.
This precludes vorticity production by the baroclinic term. The evolution equations for $\rho$ and the velocity $\uu$ are then given by
\begin{equation}
\frac{\DD\ln\rho}{\DD t}=-\nab\cdot\uu, \quad {\rm and}
\label{DlrDt}
\end{equation}
\begin{equation}
\frac{\DD\uu}{\DD t}=-\cs^2\nab\ln\rho
+\frac{\JJ\times\BB}{\rho}+\FF_\mathrm{visc},
\label{DuDt}
\end{equation}
where $\BB$ is the magnetic field, $\JJ=\nab\times\BB/\mu_0$ is the current density with $\mu_0$ being the vacuum permeability, $\JJ\times\BB$ is the Lorentz force, $\FF_\mathrm{visc}=\rho^{-1}\nab\cdot(2\nu\rho\SSSS)$ is the viscous force per unit mass with $\nu$ being the kinematic viscosity, and $\SSSS$ the rate-of-strain tensor with components ${\sf S}_{ij}= \half(\partial_i u_j+\partial_j u_i) -\onethird\delta_{ij}\nab\cdot\uu$.
Note that our simulations only include the usual shear viscosity and assume that the bulk viscosity is absent; see \cite{Beattie+23} for a recent work on this aspect.

In simulations in which the magnetic field is included, we also solve for the magnetic potential $\AAA$ via
\begin{equation}
\frac{\partial\AAA}{\partial t}=\iota\uu\times\BB+\eta\nabla^2\AAA,
\label{dAdt}
\end{equation}
so that $\nab\times\AAA$ is always divergence free.
In several of our models, we also impose an external magnetic field
$\BB_0$ by writing $\BB=\BB_0+\nab\times\AAA$, so that we can adopt
periodic boundary conditions on $\AAA$.
In \Eq{dAdt}, the parameter $\iota$ is introduced to allow us to
turn off the induction term ($\iota=0$).
By default, we have $\iota=1$.

\subsection{Vorticity Production} \label{sec:Vorticity-Production}

To understand the terms leading to vorticity production, we take the curl of \Eq{DuDt} and find
\begin{equation}
\frac{\partial\ww}{\partial t}=\nab\times(\uu\times\ww)
+\dot{\ww}_\mathrm{mag}+\dot{\ww}_\mathrm{visc},
\end{equation}
where $\dot{\ww}_\mathrm{mag}=\nab\times(\JJ\times\BB/\rho)$
is the magnetically produced vorticity and
$\dot{\ww}_\mathrm{visc}=\nab\times\FF_\mathrm{visc}$ is the
viscously produced vorticity.
Under the assumption that $\nu=\const$, we find \citep{MB06}
\begin{equation}
\dot{\ww}_\mathrm{visc}=\nu\nabla^2\ww+\nu\nab\times\GG,
\label{wdot1}
\end{equation}
where $G_i=2{\sf S}_{ij}\nabla_j\ln\rho$ is a term that always drives vorticity---even if it is initially absent.
Alternatively, if $\mu\equiv\nu\rho=\const$, we have $\FF_\mathrm{visc}=\rho^{-1}\mu(\nabla^2\uu+\onethird\nab\nab\cdot\uu)$ and
\begin{equation}
\dot{\ww}_\mathrm{visc}=\frac{\mu}{\rho}\left[\nabla^2\ww
+\nab\ln\rho\times\left(\nab\times\ww
-\fourthird\nab\nab\cdot\uu\right)\right],
\label{wdot2}
\end{equation}
when $\mu=\const$. This expression shows that viscous vorticity production results from the obliqueness of density and velocity divergence gradients, which is somewhat analogous to vorticity production by a baroclinic term in the nonisothermal case.  The $1/\rho$ term in the expression for $\dot{\ww}_\mathrm{mag}$ is generally only of minor importance when the Mach number is small. Thus, in the following, we focus on the case $\nu=\const$, where vorticity production occurs through similar terms as in the case $\mu=\const$.

While $\dot{\ww}_\mathrm{visc}$ can play a role at small scales,  it is not the only term that can convert acoustic
motions into vortical motions in a magnetized flow. This is because acoustic flows modify the magnetic field, which may then exert a Lorentz force with a finite curl. We refer to this as magnetically assisted vorticity production. We give a simple one-dimensional example of this process in \Sec{sec:CK3}, and in \Sec{sec:CK4} 
we present a set of simulations that validate the scaling relations obtained from the one-dimensional model.

\subsection{Initial velocity field} \label{sec:Summary_runs}

Our study is based on the {\sc Pencil Code} \citep{PC}, which employs sixth-order centered differences and a third-order time-stepping scheme. In all cases, we use a resolution of $1024^3$ mesh points. Our simulations have periodic boundary conditions, so the mass in the volume is conserved, and the mean density $\rho_0\equiv\bra{\rho}$ is constant. Here and below, angle brackets denote volume averaging.

Our initial velocity and vector potential are constructed in Fourier space as
$\uu(\xx)=\sum\tilde{\uu}(\kk)\,e^{\ii\kk\cdot\xx}$ and
$\AAA(\xx)=\sum\tilde{\AAA}(\kk)\,e^{\ii\kk\cdot\xx}$ with
\begin{equation}
\tilde{u}_i(\kk)=\left[(1-\zeta)\delta_{ij}-(1-2\zeta)\hat{k}_i\hat{k}_j\right]\,u_\mathrm{ini}\tilde{S}_j(\kk),
\end{equation}
\begin{equation}
\tilde{A}_i(\kk)=\left(\delta_{ij}-\hat{k}_i\hat{k}_j\right)\,A_\mathrm{ini}\tilde{S}_j(\kk).
\end{equation}
Here, $u_\mathrm{ini}$ and $A_\mathrm{ini}$ are amplitude factors, $\hat{k}_i$ are the
components of the unit vector $\hatkk\equiv\kk/k$, $\tilde{S}_j(\kk)$
is a vector field in Fourier space with three independent components
that depend on $k=|\kk|$ but have random phases $\varphi(\kk)$ for each
$\kk$ vector, and $\zeta$ is the irrotationality parameter with
$\zeta=0$ when the initial velocity is vortical and $\zeta=1$
when it is acoustic (irrotational).
Here, we choose
\begin{equation}
\tilde{S}_j(\kk)=\frac{k_0^{-3/2} (k/k_0)^{\alpha/2-1}}
{1+(k/k_0)^{(\alpha+5/3)/2}}\,e^{\ii\varphi(\kk)},
\end{equation}
where $k_0$ is the peak wavenumber of the initial condition,
and $\alpha$ is the slope of the subinertial range, which we set to $\alpha=4$ in this work.

\begin{table*}\caption{
Summary of the runs discussed in this paper.
Columns show 
the run name (column~1),
irrotationality parameter $\zeta$ (column~2),
induction switch $\iota$ (column~3),
the normalized peak wavenumber $\tilde{k}_0=k_0/k_1$ (column~4),
the normalized amplitudes of initial random velocity and vector potential,
$\tilde{u}_\mathrm{ini}=u_\mathrm{ini}/\cs$ and $\tilde{A}_\mathrm{ini}=k_1 A_\mathrm{ini}/\sqrt{\rho_0\mu_0}\cs$ 
(columns~5 and 6),
five different Mach numbers (columns~7--11),
the Reynolds number (column~12),
and six different Kolmogorov-type parameters (columns~13--18).
Dashes indicate that $C_\mathrm{M}$ cannot be determined for nonmagnetic runs.
Run~D is the same as Run~C, except that the induction term has been ignored in \Eq{dAdt}.
}\hspace{-22mm}\vspace{12pt}\centerline{\begin{tabular}{cccc ccc | cccc cccc ccc}
\multicolumn{7}{c}{input parameters} & \multicolumn{11}{c}{output parameters} \\
Run & $\zeta$ & $\iota$ & $\tilde{k}_0$ & $\tilde{u}_\mathrm{ini}$ & $\tilde{A}_\mathrm{ini}$ &
$\mathrm{Ma}_\mathrm{M0}$ &
$\mathrm{Ma}_\mathrm{M1}$ &
$\mathrm{Ma}_\mathrm{K}$ &
$\mathrm{Ma}_\mathrm{V}$ &
$\mathrm{Ma}_\mathrm{A}$ &
$\Rey$ &
$C_\mathrm{M}$ &
$C_\mathrm{K}$ &
$C_\mathrm{KV}$ &
$C_\mathrm{KA}$ &
$C_\mathrm{V}$ &
$C_\mathrm{A}$ \\
\hline
A &0 & 1 & 10 & 0.020 &   0   &  0   &   0   & 0.020 & 0.020 & 0.002 &  1200 &  ---  &  1.62 &  1.62 &  0.00 &  1.65 &  0.00 \\
B &1 & 1 & 10 & 0.020 &   0   &  0   &   0   & 0.013 & 0.000 & 0.013 &  1000 &  ---  &  6.06 &  0.00 &  6.06 &  0.35 &  6.06 \\
C &1 & 1 & 10 & 0.020 & 0.005 &  0   & 0.009 & 0.014 & 0.004 & 0.013 &  1100 &  2.93 &  6.31 &  0.33 &  5.99 &  0.80 &  7.22 \\
D &1 & 0 & 10 & 0.020 & 0.005 &  0   & 0.019 & 0.033 & 0.031 & 0.010 &  1200 &  0.00 &  1.86 &  1.65 &  0.22 &  1.69 &  0.48 \\
E &0 & 1 & 10 &   0   & 0.005 &  0   & 0.008 & 0.003 & 0.003 & 0.000 &   200 &  2.88 &  0.96 &  0.96 &  0.00 &  0.96 &  0.02 \\
F &1 & 1 & 10 & 0.020 &   0   & 1.00 & 0.010 & 0.014 & 0.008 & 0.012 &   200 &  3.66 &  3.92 &  1.57 &  2.35 &  3.09 &  3.17 \\
G &1 & 1 &  2 & 0.020 & 0.005 &  0   & 0.014 & 0.013 & 0.007 & 0.011 &  2100 &  1.80 &  6.58 &  0.86 &  5.72 &  1.54 &  7.36 \\
H &1 & 1 &  2 & 0.020 &   0   &  0   &   0   & 0.026 & 0.000 & 0.026 &  1300 &  ---  &  2.17 &  0.00 &  2.17 &  0.26 &  2.18 \\
I &1 & 1 &  2 & 0.020 &   0   & 0.02 &   0   & 0.026 & 0.000 & 0.026 &  1900 &  0.19 &  2.26 &  0.00 &  2.26 &  0.18 &  2.27 \\
J &1 & 1 &  2 & 0.020 &   0   & 0.05 & 0.001 & 0.026 & 0.000 & 0.026 &  1900 &  0.35 &  2.26 &  0.00 &  2.26 &  0.32 &  2.28 \\
K &1 & 1 &  2 & 0.020 &   0   & 0.10 & 0.002 & 0.026 & 0.001 & 0.026 &  1500 &  0.57 &  2.13 &  0.00 &  2.12 &  0.74 &  2.13 \\
L &1 & 1 &  2 & 0.020 &   0   & 0.20 & 0.004 & 0.026 & 0.001 & 0.026 &  1300 &  0.98 &  2.11 &  0.02 &  2.09 &  1.39 &  2.09 \\
M &1 & 1 &  2 & 0.020 &   0   & 0.50 & 0.011 & 0.026 & 0.005 & 0.025 &  1900 &  1.95 &  2.47 &  0.17 &  2.30 &  2.18 &  2.35 \\
N &1 & 1 &  2 & 0.020 &   0   & 1.00 & 0.020 & 0.028 & 0.015 & 0.023 &  1900 &  2.92 &  3.13 &  1.01 &  2.12 &  2.32 &  2.67 \\
O &1 & 1 &  2 & 0.004 &   0   & 0.10 & 0.001 & 0.008 & 0.000 & 0.008 &   500 &  0.63 &  2.74 &  0.00 &  2.74 &  0.16 &  2.76 \\
P &1 & 1 &  2 & 0.004 &   0   & 1.00 & 0.006 & 0.008 & 0.004 & 0.007 &   500 &  2.90 &  3.12 &  0.55 &  2.57 &  1.99 &  2.87 \\
\label{TSummary}\end{tabular}}\end{table*}

\subsection{Diagnostic Quantities} \label{sec:Diagnostic}

An important characteristic of turbulence is its energy spectrum.
The kinetic energy density per linear wavenumber interval, $\EK(k,t)$, is defined
as the modulus squared of the Fourier transform of the velocity integrated
over concentric shells in wavevector space. The spectrum is normalized such that $\EEK(t)=\int\EK(k,t)\,\dd k$ is the mean kinetic energy density.
To obtain the energy per unit volume, we include a $\rho_0$ factor,
so $\EEK(t)=\rho_0\bra{\uu^2}/2$, but we refer the reader to \cite{Kritsuk+07}
for alternatives.

The magnetic energy spectrum $\EM(k,t)$ is defined analogously such that
$\EEM(t)=\int\EM(k,t)\,\dd k$ is the mean magnetic energy density with
$\EEM(t)=\bra{\BB^2}/2\mu_0$.
In addition, we also compute the spectrum of the vorticity, $E_w(k,t)$,
analogously to $\EK(k,t)$, but with the velocity $\uu$ being replaced
by the vorticity $\ww=\nab\times\uu$.
In this case, $E_w(k,t)$ is related to the vortical part of
the kinetic energy spectrum, $E_\mathrm{V}(k,t)$, through
$E_\mathrm{V}(k,t)=E_w(k,t)/k^2$.

Finally, we also consider the scaled logarithmic density spectrum,
$E_{\ln\!\rho}(k,t)$, which is normalized such that
$\int E_{\ln\!\rho}(k,t)\,\dd k=\rho_0\bra{(\cs\ln\!\rho)^2}/2$.
Looking at \Eq{DlrDt}, the spatiotemporal
Fourier transform of its linearized form reads
$-\ii\omega\,\widetilde{\ln\rho}=-\ii\kk\cdot\tilde{\uu}$,
where $\omega$ is the frequency.
Using the dispersion relation for sound waves, $\omega=\cs k$, we have
$\cs\widetilde{\ln\rho}=\hat{\kk}\cdot\tilde{\uu}$, so that $\cs\ln\rho$ is
directly a proxy for the longitudinal velocity, and
$E_{\ln\!\rho}(k,t)$ is a proxy of the energy spectrum of the
acoustic part, $E_\mathrm{A}\approx E_{\ln\!\rho}$.
We note that $\EK=E_\mathrm{V}+E_\mathrm{A}$ is a fairly accurate decomposition, at least for subsonic flows.
We therefore compute the acoustic velocity spectrum as
$E_\mathrm{A}=E_\mathrm{K}-E_\mathrm{V}$, and have verified that
$E_{\ln\!\rho}$ is indeed a good approximation of $E_\mathrm{A}$.

The kinetic and magnetic dissipation rates are
\begin{equation}
\epsK\equiv\bra{2\nu\rho\SSSS^2},\qquad {\rm and} \qquad \epsM\equiv\bra{\eta\mu_0\JJ^2},
\end{equation}
respectively.
The magnetic dissipation can also be obtained from $\epsM(t)=\int2\eta k^2\EM(k,t)\,\dd k$.
For the kinetic energy dissipation, however, we have to remember that
vortical and irrotational parts contribute differently, because
\begin{equation}
\bra{\SSSS^2}=\bra{\ww^2}+\fourthird\!\bbra{(\nab\cdot\uu)^2}.
\end{equation}
Therefore, we also define $\epsV(t)=\int2\nu k^2\EV(k,t)\,\dd k$,
and $\epsA(t)=\fourthird\int2\nu k^2\EA(k,t)\,\dd k$, but note that,
in general, $\epsK\neq\epsV+\epsA$ owing to the existence of mixed terms.

To characterize the velocity and magnetic fields of our runs, we define
five different Mach numbers. The usual Mach number is $\Ma=\urms/\cs$, which characterizes the combined vortical and acoustic parts. This can also be characterized
separately through $\Ma_\mathrm{V}=\sqrt{2\EEV/\rho_0}/\cs$
and $\Ma_\mathrm{A}=\sqrt{2\EEA/\rho_0}/\cs$, so that
$\Ma^2=\Ma_\mathrm{V}^2+\Ma_\mathrm{A}^2$
The magnetic field is characterized by the Alfv\'en speed
$\vA=\Brms/\sqrt{\rho_0\mu_0}$, which allows us to define a corresponding
Mach number. Here, it is convenient to consider separately the contributions from
the imposed field $\vAz=\BB_0/\sqrt{\rho_0\mu_0}$ and the rest, $\vAo$,
so that $\vA^2=\vAz^2+\vAo^2$. The corresponding Mach numbers are then $\Ma_\mathrm{M0}=\vAz/\cs$ and $\Ma_\mathrm{M1}=\vAo/\cs$.

We also define the time-dependent Reynolds number $\Rey=\urms\xiK/\nu$ based on the usual integral scale:
\begin{equation}
\xiK=\int k^{-1}\EK(k)\,dk/\EEK
\end{equation}
and quote in the following a late-time average when it varies only slowly.
In all cases, our Mach numbers are averaged over a fixed interval at a time
of around one hundred sound travel times, $(\cs k_1)^{-1}$.
The values of the Mach numbers are well below unity, and
the magnetic Prandtl number, $\Pm=\nu/\eta$, is taken to be unity in all cases.

It is convenient to present magnetic and kinetic energy spectra in normalized form.
Instead of normalizing them by a quantity characterizing the large-scale
properties ($\EEK/k_0$), we choose here to normalize them by the quantity
$\epsK^{2/3}/k_0^{5/3}$ characterizing the small scales.
For our runs, we take the values $k_0/k_1=10$ and $2$.

\section{Results} \label{sec:results}

\subsection{Summary of the Runs}

We have performed a series of runs varying the input parameters
$\zeta$, $\iota$, $k_0$, $u_\mathrm{ini}$, $A_\mathrm{ini}$, $\mathrm{Ma}_\mathrm{M0}$, and
$\mathrm{Ma}_\mathrm{M1}$; see \Tab{TSummary} for a summary.
Nonmagnetic runs are those where
$\mathrm{Ma}_\mathrm{M0}=\mathrm{Ma}_\mathrm{M1}=0$ (see Runs~A, B, and H).
When $\mathrm{Ma}_\mathrm{M0}=0$, but $\mathrm{Ma}_\mathrm{M1}\neq0$, we have initially a random (`turbulent') magnetic
field with a spectrum peaking at $k\approx k_0$, similar to the
initial velocity field (Runs~C--E and G).
Run~D is the same as Run~C, except that the induction term has been ignored in \Eq{dAdt}, i.e., $\iota=0$.
The Mach numbers are in the range 0.007--0.04, and the Reynolds number is in the range 200--1900.

The Kolmogorov-type parameter or prefactor for the magnetic field, $C_\mathrm{M}$, varies significantly and is usually in the range 2--8. In all cases with $\zeta\neq0$, $C_\mathrm{K}$ exceeds the typical value
of 1.6 for vortical turbulence (Run~A).
Almost no vorticity is produced when $\mathrm{Ma}_\mathrm{K}$ and
$C_\mathrm{KV}$ are small (Runs~B, C, H--L, and O).
This is the case for all nonmagnetic and weakly magnetized cases when $\mathrm{Ma}_\mathrm{M0}\la0.1$.
Vorticity is produced when $\mathrm{Ma}_\mathrm{M0}\ga 0.01$ or $\mathrm{Ma}_\mathrm{M1}\ga0.02$ (Runs~C--E and G).
We recall that Run~G has a smaller value of $k_0$ than Runs~C--E.

\subsection{Comparison of Typical Spectra}

\begin{figure}\begin{center}
\includegraphics[width=\columnwidth]{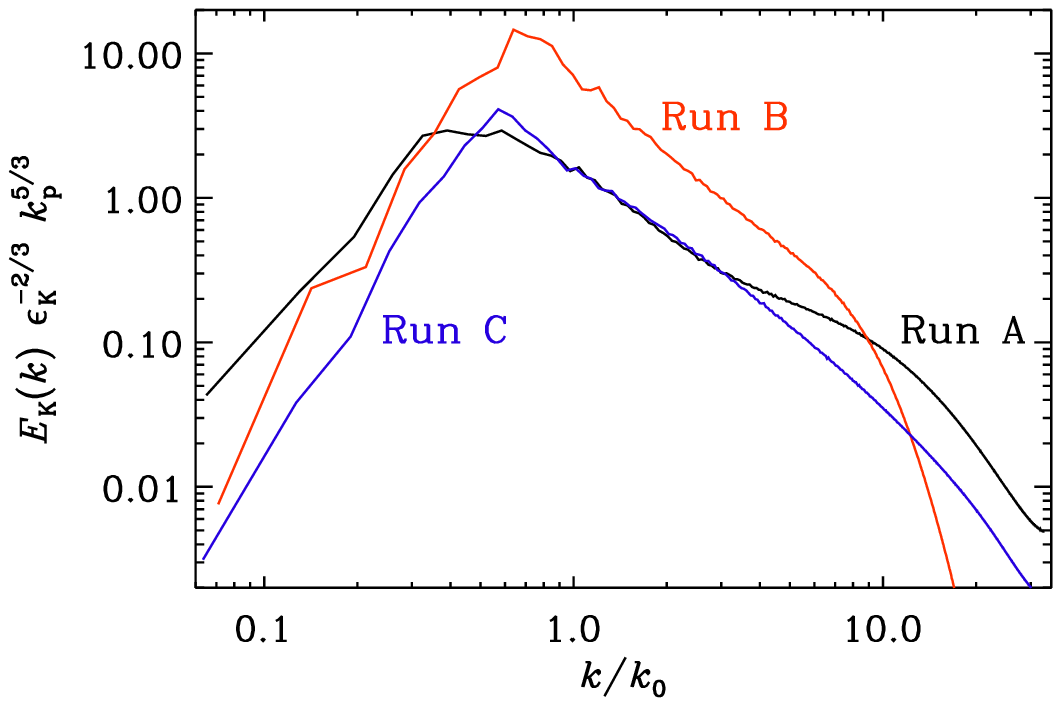}
\end{center}\caption{Kinetic energy spectra for Runs~V (black), A (red), and C (blue), all at time $t=28/\cs k_1$. No distinction between vortical and acoustic contributions has been made.}\label{pspec_comp}\end{figure}

The velocity spectra for Run~A, with vortical hydrodynamic turbulence, Run~B, with acoustic hydrodynamic turbulence, and Run~C, with acoustic MHD turbulence,
are compared at a fixed time in \Fig{pspec_comp}.
We see that, although our runs have a fixed viscosity ($\nu k_1/\cs=10^{-6}$ for $k_0/k_1=10$ and $\nu k_1/\cs=5\times10^{-6}$ for $k_0/k_1=2$),
and similar values for the Mach Number, only Run~A has a spectrum that still possesses significant energy at large $k$.
It is also the only run with a marked bottleneck, i.e., a shallow part just before the viscous subrange at large $k$ \citep{Falk94}.
The peak of the scaled spectra for Run~B is higher, reflecting the fact that the Kolmogorov prefactor for acoustic turbulence is larger, as we discuss below.
Finally, the kinetic energy spectra for Runs~B and C are similar to that for Run~A, except there is no visible bottleneck.

\subsection{Kolmogorov Prefactor} \label{sec:CK}

In Kolmogorov's theory, the constancy of the kinetic energy flux along the turbulent cascade makes $\epsK$ an important quantity for dimensional arguments. On dimensional grounds, the spectrum should be equal to $C_\mathrm{K}\epsK^{2/3}k^{-5/3}$, where $C_\mathrm{K}$, the dimensionless prefactor, is the Kolmogorov constant \citep{Frisch95}. To obtain the value of $C_\mathrm{K}$, it is convenient to present compensated spectra, $\epsK^{-2/3}k^{5/3}\EK(k,t)$, which should show a constant plateau in the $k$ range where the Kolmogorov scaling applies. Note that the difference with our normalization in \Fig{pspec_comp} lies in the fact that there the factor $k_0^{5/3}$ was a constant, but now it is $k$ dependent.

\begin{figure}\begin{center}
\includegraphics[width=\columnwidth]{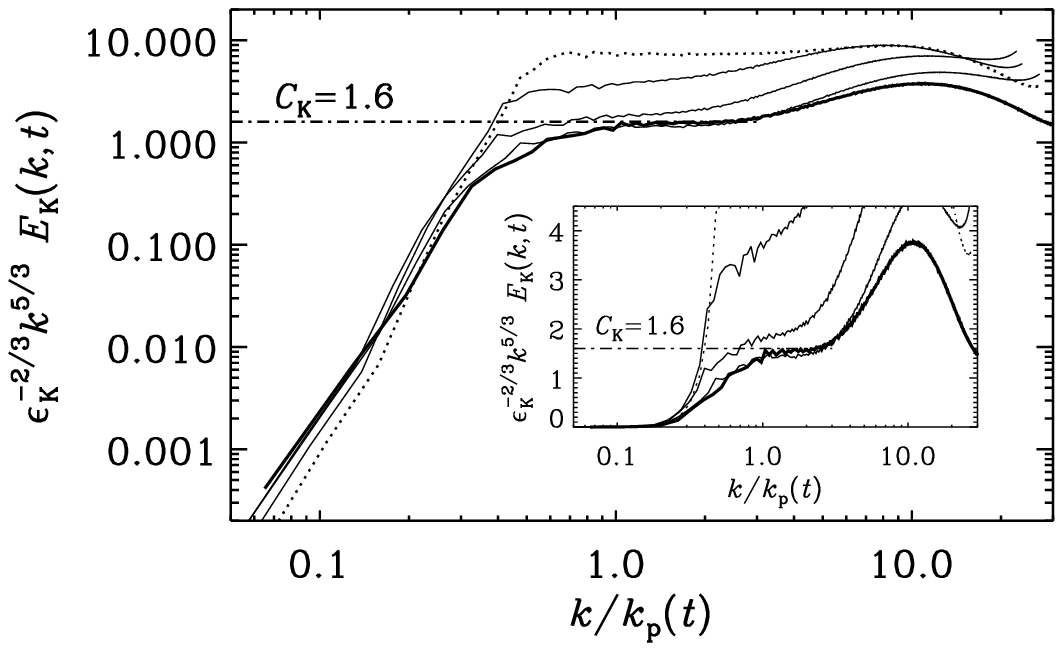}
\end{center}\caption{ Compensated kinetic energy spectra for Run~A at times $\cs k_1 t=3$, 7, 14, and 28. The dotted line denotes the initial state, and the thick line marks the last time. The dashed--dotted horizontal line marks the approach to the value $C_\mathrm{K}=1.5$. The inset shows the approach to a plateau in a semilogarithmic plot.}\label{ppower_Vrun}
\end{figure}

We begin with the more familiar vortical case with $\zeta=0$ and no magnetic field ($\BB=0$, Run~A). The result is shown in \Fig{ppower_Vrun}, where we see the approach to a plateau in the compensated spectrum at the level $\CK\approx1.6$, which agrees with the usual Kolmogorov constant \citep{Kaneda+03, Bra+23}. Near the dissipative subrange, we also see a strong bulge. This was already evident from \Fig{pspec_comp} and was characterized as the bottleneck \citep{Falk94}. It is significantly stronger here than for ordinary (stationary) turbulence, for which the compensated spectrum at the bottleneck is usually well below 3 \citep{Kaneda+03, HBD04, Bra+23}. This could partially be a consequence of having underresolved the high wavenumbers at early times.

\begin{figure}\begin{center}
\includegraphics[width=\columnwidth]{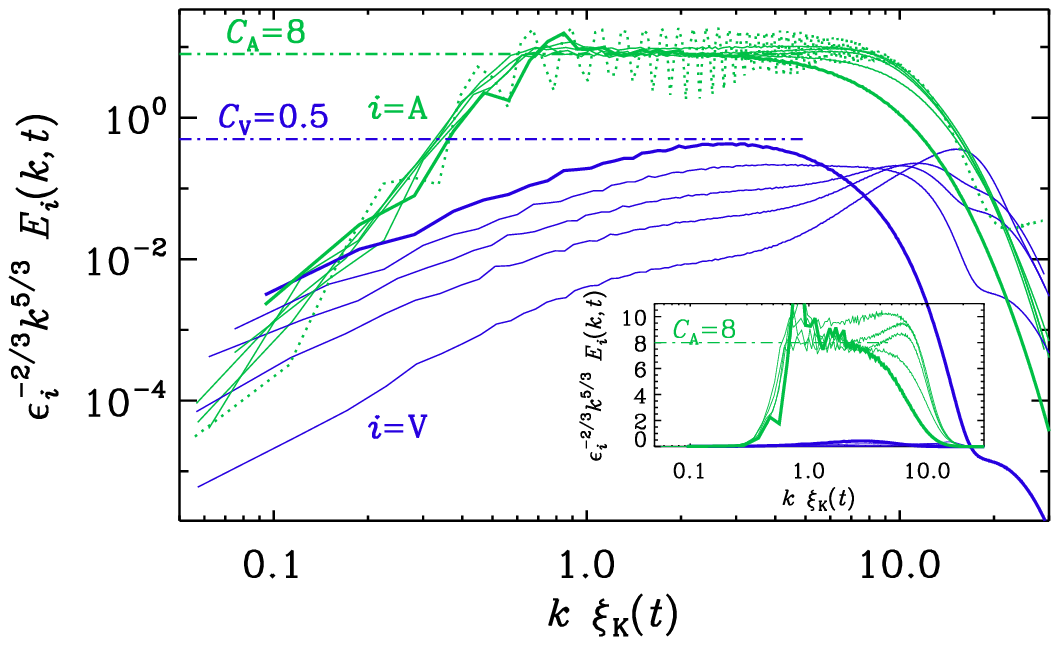}
\end{center}\caption{Compensated kinetic energy spectra for acoustic turbulence (Run~B), $\epsilon_i^{-2/3}(t)\,k^{5/3}E_i(k,t),$ separated into the vortical ($i=\mathrm{V}$, blue lines) and acoustic ($i=\mathrm{A}$, green lines) components. }\label{ppower_Arun}\end{figure}

The corresponding case for acoustic turbulence, where $\zeta=1$, looks different in many ways. This is shown in \Fig{ppower_Arun}, where we plot spectra that are compensated separately for the vortical and acoustic parts, i.e.,
\begin{equation}
c_i(k,t)=\epsilon_i^{-2/3}(t)\,k^{5/3}E_i(k,t),
\label{ci-spectrum}
\end{equation}
and denote by $C_i$ the approximate average of $c_i(k,t)$ over the flat part for $i=$ V or A. Here, we still see the approach to a plateau, but the bottleneck is very weak (see the inset). Instead, there is a spike in $E_\mathrm{A}$ at the low-wavenumber end, where the spectrum transits from the subinertial range to the inertial range. In the following, we refer to this spike as the subinertial range peak. The height of the plateau also significantly exceeds the usual value for vortical turbulence and is $\CA\approx8$, suggesting that the standard Kolmogorov phenomenology may not be applicable.

In Run~B, some vorticity is produced by the interaction with viscosity. Although the spectrum in \Fig{ppower_Arun} is normalized by $\epsA$, the level of the plateau is low (around 0.5), although it still increases over time. As discussed in \Sec{sec:equations}, this vorticity production results from the obliqueness of density and velocity divergence gradients. We find that the amount of vorticity production is virtually the same regardless of whether $\nu$ or $\mu$ is constant.  This is likely because the Mach number is small in both cases, meaning that density fluctuations are also small. 

\subsection{Magnetic Vorticity Production} \label{sec:CK}

Next, for Run~C, we consider an irrotational initial flow ($\zeta=1$, just like Run~B) together with a random initial magnetic field
with a spectrum $\EM\propto k^4$ for $k<k_0$ and $\EM\propto k^{-5/3}$ for $k>k_0$, just like the initial velocity field.
Depending on the relative strengths of the magnetic and velocity fields, the curl of the Lorentz force can drive vorticity through the $\dot{\ww}_\mathrm{visc}$ term in \Eqs{wdot1}{wdot2}.

\begin{figure}\begin{center}
\includegraphics[width=\columnwidth]{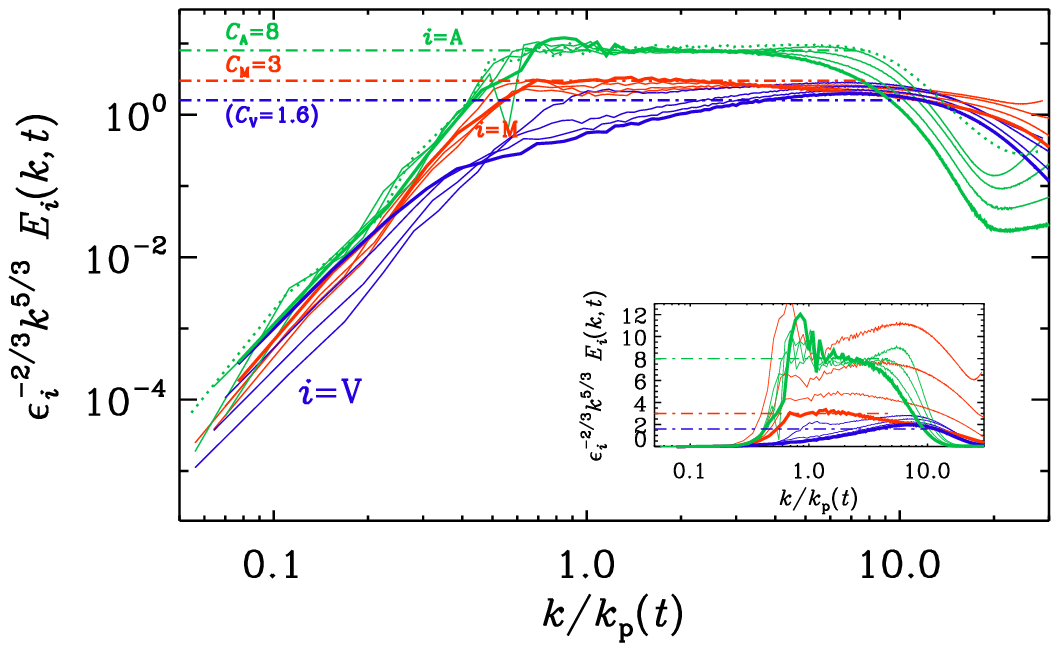}
\end{center}\caption{ Similar to \Fig{ppower_Arun}, but for Run~C, where the magnetic field produces vorticity. Compensated magnetic energy spectra are also plotted ($i=\mathrm{M}$, red lines). The dashed--dotted horizontal lines indicate the approximate positions of plateaus at $C_\mathrm{A}\approx8$ (green), $C_\mathrm{M}=3$ (red), and $C_\mathrm{V}=2$ (blue). }\label{ppower_M1run}\end{figure}

The result for Run~C is shown in \Fig{ppower_M1run}.
Interestingly, the magnetic energy $\EM(k)$ shows neither a marked bottleneck nor a marked subinertial range peak. The compensated $c_\mathrm{V}(k)$ spectrum of \Eq{ci-spectrum} does not have a plateau, but it crosses $\CV\approx1.6$ at intermediate wavenumbers. Note, however, that $\CM(t)$ has a plateau with a magnetic Kolmogorov prefactor of about 3; see \Fig{ppower_M1run}.

In the Appendix, we compare spectra for Runs~C and E with and without initial turbulence, Runs~C and D with and without the induction term, i.e., $\iota=1$ and 0, respectively, as well as Runs~C and G with $k_0/k_1=10$ and 2, respectively.
We see that in Run~E, turbulence is gradually produced.
Regarding the presence or absence of the induction term, we see that for Run~C the induction term enables the magnetic and kinetic energy cascades to be nearly parallel.
This is not the case when the induction term is absent (Run~D).
Finally, comparing Runs~C and G, we see that for both $k/k_1=2$ and 10, there is a loss of kinetic energy in the acoustic components along with a gain of kinetic energy in the vortical component.

\subsection{Comparison with Earlier Work} \label{sec:earlier}

\begin{figure}
\includegraphics[width=\columnwidth]{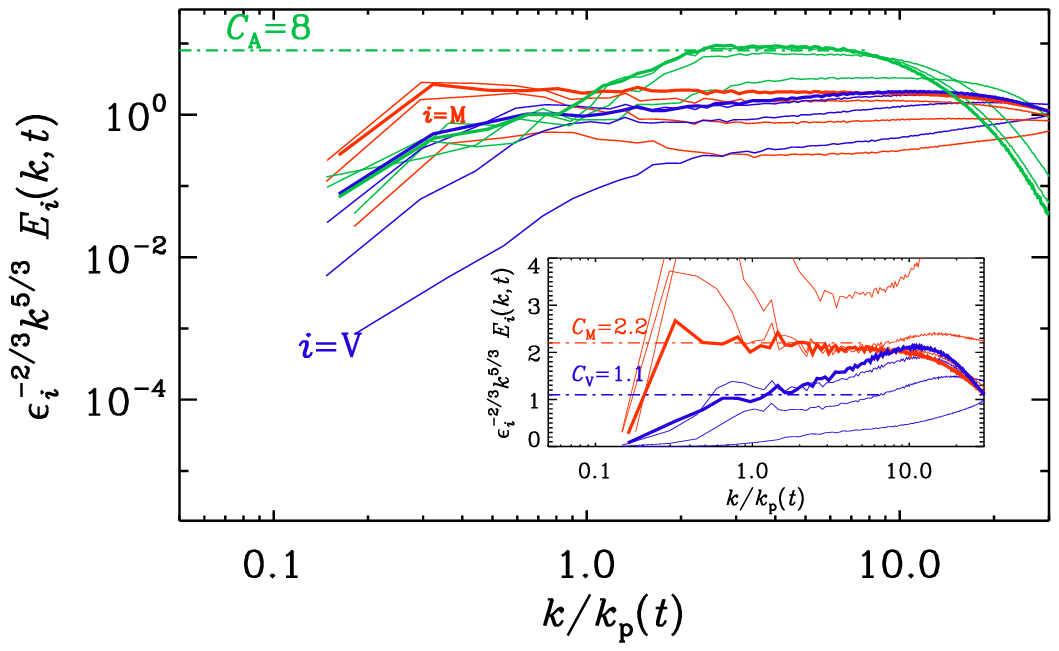}
\caption{The same as \Fig{ppower_M1run}, but for Run~G
with $k_0/k_1=2$.
\label{ppower_MO2run}}
\end{figure}

In the presence of irrotational forcing, \cite{Kahn+12} found that, for an inflationary magnetic field with a magnetic energy spectrum proportional to $k^{-1}$, vortical turbulence develops with a spectrum $\EV(k)$ that is in equipartition, i.e., $\EV(k)\approx\EM(k)$. Comparing this with our present results, we see that equipartition between $\EV(k)$ and $\EM(k)$ exists only at high wavenumbers. This difference with \cite{Kahn+12} seems to be connected with the fact that they used an inflationary magnetic field with a $k^{-1}$ spectrum, whereas here $\EM(k)$ has a peak at intermediate wavenumbers. To further verify this interpretation, we show in \Fig{ppower_MO2run} the compensated spectra of $\EV(k,t)$ and $\EM(k,t)$ for a run with $k_0/k_1=2$. Here, we see that the range over which the two spectra are nearly parallel is not only increased, but also the degree of equipartition is better, i.e., the two spectra are closer together.

The velocity spectrum generated by the Lorentz force of such a magnetic field alone, i.e., without an initial acoustic component, is known to develop a shallow spectrum near $k_0$, and is in approximate equipartition with the magnetic field at large wavenumbers. This is similar to the $\EV$ spectrum in \Fig{ppower_M1run}, where the compensated spectra are proportional to $k^{2/3}$, suggesting that $\EV(k)\propto k$ in the beginning of the magnetic inertial range.

In agreement with the earlier work of \cite{MB06}, the present results confirm that acoustic turbulence hardly contributes to driving magnetic fields. Theoretically, small-scale dynamo action of the type first proposed by \cite{Kaz68} should also be possible for acoustic turbulence \citep{Kazantsev+85, Afonso+19}, but this has never been confirmed numerically \citep{MB06}. What has been confirmed, however, is a small negative turbulent magnetic diffusivity \citep{Radler+11}. Because its negative value is never larger than the positive microphysical magnetic diffusivity, it can only slow down the decay without leading to dynamo action from this effect alone. Furthermore, this negative turbulent magnetic diffusivity effect only concerns the mean or large-scale magnetic field.

\subsection{Magnetically Assisted Vorticity Conversion} \label{sec:CK2}

As we have seen from \Tab{TSummary}, runs with sufficiently strong
uniform magnetic fields produce noticeable amounts of vorticity.
Here, the mechanism causing vorticity is different from the vorticity production considered in \Sec{sec:CK}, because it relies on the presence of initially acoustic turbulence.
This is what we call magnetically assisted vorticity conversion.
To gain a better understanding of this mechanism,
we first consider a simple one-dimensional example.

\subsubsection{Vorticity Conversion in One Dimension} \label{sec:CK3}

\begin{figure}
\includegraphics[width=\columnwidth]{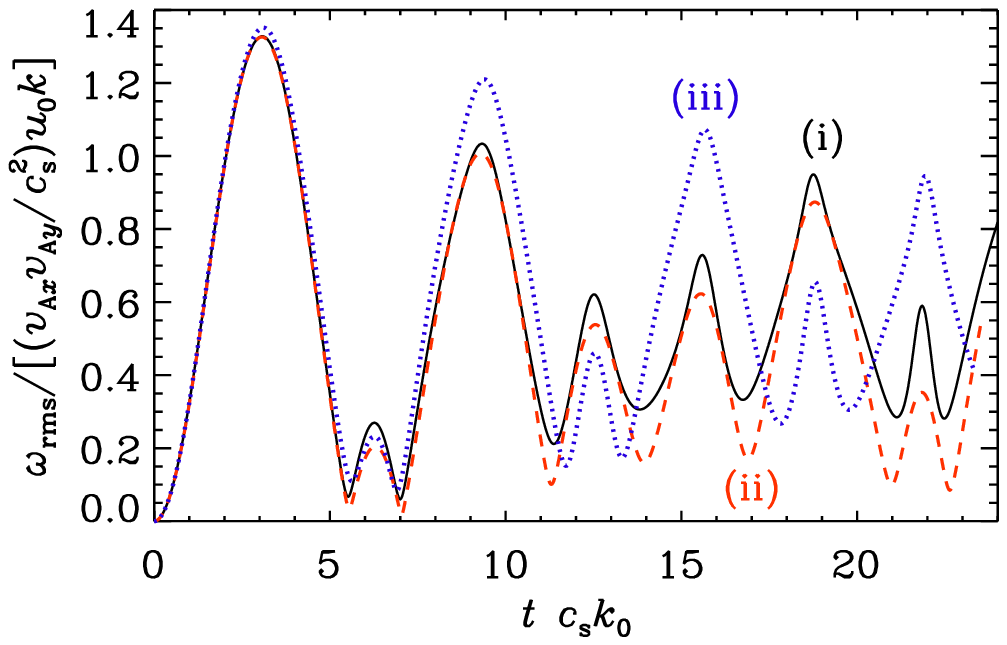}
\caption{
(i) $u_0=0.1$, $B_0=0.1$;
(ii) $u_0=0.05$, $B_0=0.1$;
(iii) $u_0=0.1$, $B_0=0.05$.
Note that the normalized curves of $w_{\rm rms}$ for all three cases are initially the same.
\label{pcomp}}
\end{figure}

The conversion of acoustic kinetic energy into vortical kinetic energy
can be demonstrated with the help of a one-dimensional example.
We consider a domain $-\pi<x<\pi$ with a uniform magnetic field in
the diagonal direction, $\BB_0=(B_{0x},B_{0y},0)$, constant density,
$\rho=\rho_0$, and a standing sound wave initially, i.e., $u_x=u_0\sin kx$.
All the kinetic energy is in acoustic motions.
The uncurled induction equation reads $\dot{A}_z=u_x B_{0y}$, and
the momentum equation becomes $\dot{\uu}=J_z (-B_{0y},B_{0x},0)/\rho_0$,
where dots denote time derivatives.
For vorticity production, which yields $w_z=u'_y$, where primes denote $x$ derivatives, only the $u_y$ component matters, and thus we have $\dot{w}_z=J_z' B_{0x}/\rho_0$,
Taking another time derivative and using $J_z=-A_z''$, we have
$\ddot{w}_z=-u_x'''v_{\mathrm{A}x}v_{\mathrm{A}y}$, where
$v_{\mathrm{A}x}$ and $v_{\mathrm{A}y}$ are the Alfv\'en speeds in
the $x$ and $y$ directions, respectively.
Replacing $t$ and $x$ derivatives by factors of $\omega$ and $k$, and
using the dispersion relation for sound waves, $\omega=\cs k$, we find
for the vorticity amplitude
\begin{equation}
w_z=(v_{\mathrm{A}x}v_{\mathrm{A}y}/\cs^2)\,u_0 k.
\label{wz-scaling}
\end{equation}
For $u_x=u_0\sin kx$, $w_z$ is proportional to $\cos kx$.
In \Fig{pcomp}, we show three cases: (i) $u_0=0.1$, $B_0=0.1$;
(ii) $u_0=0.05$, $B_0=0.1$; and (iii) $u_0=0.1$, $B_0=0.05$.
We see that the linear scaling in $u_0$ and the quadratic scaling in
$B_0$ in \Eq{wz-scaling} is reproduced by a numerical simulation of this
one-dimensional initial value problem.

\begin{figure}
\includegraphics[width=\columnwidth]{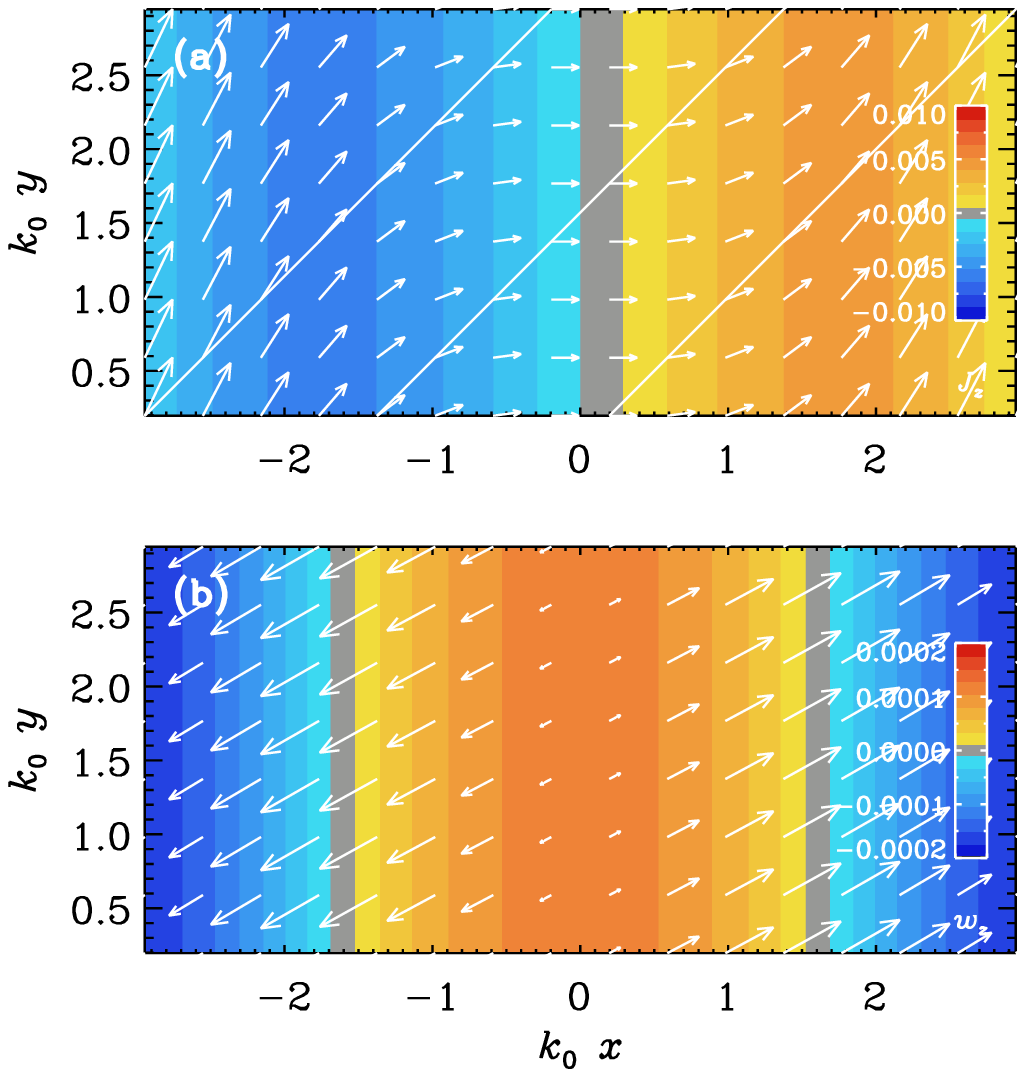}
\caption{Visualization of $(B_x,B_y)$ vectors overlaid on
a color-scale representation of $J_z$ (a) and
of $(u_x,u_y)$ vectors overlaid on $w_z$ (b)
in a two-dimensional plane by replicating the
data of the one-dimensional calculation in the $y$ direction.
\label{ppvar}}
\end{figure}

\begin{figure}
\includegraphics[width=\columnwidth]{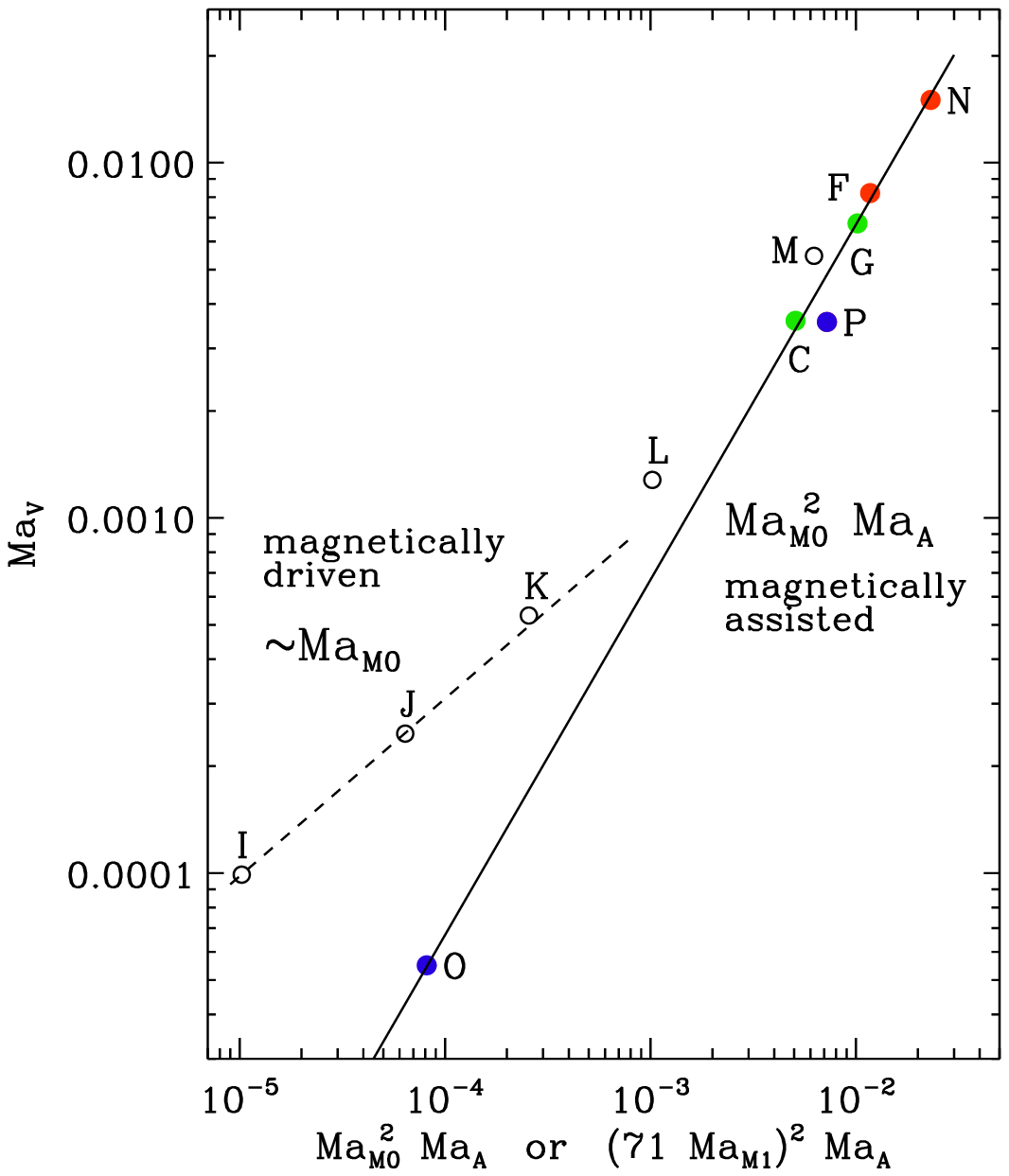}
\caption{Dependence of $\mathrm{Ma}_\mathrm{V}$ on
$\mathrm{Ma}_\mathrm{M0}^2\mathrm{Ma}_\mathrm{A}$ for our three-dimensional runs with an imposed magnetic field, and on
$(71\mathrm{Ma}_\mathrm{M1})^2\mathrm{Ma}_\mathrm{A}$ for Runs~C and G without imposed magnetic field. The red filled symbols mark Runs~F and N, while the blue filled symbols mark Runs~O and P. The green filled symbols mark Runs~C and G without an imposed magnetic field. The solid line corresponds to $0.67\,\mathrm{Ma}_\mathrm{M0}^2\mathrm{Ma}_\mathrm{A}$ and
the dashed line to $0.03\,\mathrm{Ma}_\mathrm{M0}$.
The uppercase letters denote the runs.
\label{presults_Cm_long_1panel}}
\end{figure}

\begin{figure*}
\includegraphics[width=\textwidth]{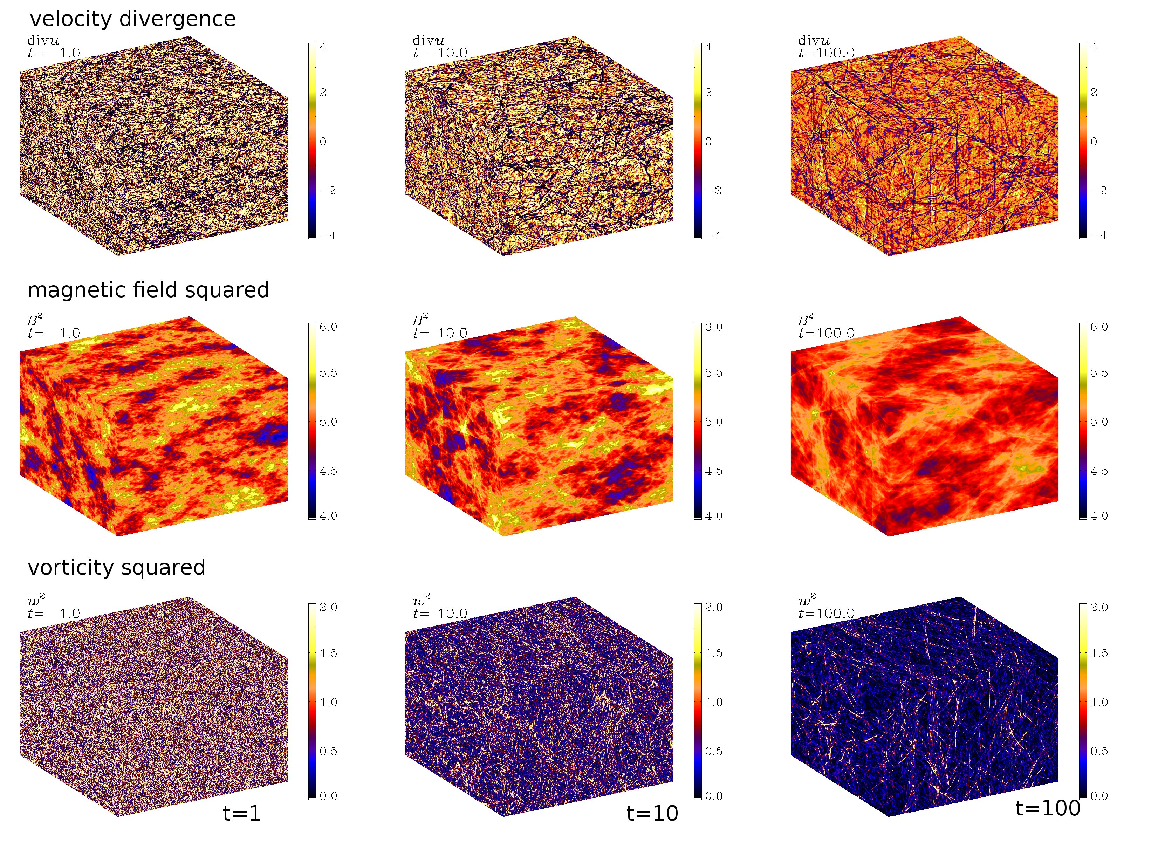}
\caption{Visualizations of $\nab\cdot\uu$ (top) with a range of $-4$ to $4$, $\BB^2$ (middle) with a range of 4--6, and $\ww^2$ (bottom) with a range of 0 to 2.  All plots are on the periphery of the computational domain for Run~N at $t\,\cs k_1=1$, 10, and 100. \label{ABC_large}}
\end{figure*}

In \Fig{ppvar} we present visualizations of $(B_x,B_y)$ vectors
and $(u_x,u_y)$ vectors overlaid on color-scale representations of $J_z$ and $w_z$, respectively. To make the small departures from the uniform field more clearly visible, we have scaled the perturbations of $B_y$ by a factor of 20 and $u_y$ by a factor of 300.

Given that the magnetically assisted conversion of acoustic into vortical motions requires strong fields, it is of interest
to see whether the strength of this conversion agrees with what is implied by \Eq{wz-scaling}. This is done in the next section.

\subsubsection{Vorticity Conversion in Three Dimensions} \label{sec:CK4}

To see if the scaling found in \Sec{sec:CK3} applies to our runs, we plot in \Fig{presults_Cm_long_1panel} the dependence of $\mathrm{Ma}_\mathrm{V}$ on
$\mathrm{Ma}_\mathrm{M0}^2\mathrm{Ma}_\mathrm{A}$ for runs with an imposed magnetic field. Except for Runs~I--L with $0.02\leq\mathrm{Ma}_\mathrm{M1}\leq0.2$, in which the magnetic
field is weak and the acoustic turbulence strong, the vorticity obeys the expected scaling with $\mathrm{Ma}_\mathrm{V}\approx0.67\,\mathrm{Ma}_\mathrm{M0}^2\mathrm{Ma}_\mathrm{A}$. For runs without an imposed magnetic field, the same scaling can also be recovered if
we multiply $\mathrm{Ma}_\mathrm{M1}$ by a factor of $\approx71$, suggesting that a much weaker turbulent field has the same effect as a stronger uniform field. Note that it is difficult to distinguish this type of conversion from vorticity produced directly from the Lorentz force (here Run~E). However, we see that the expected dependence on $\mathrm{Ma}_\mathrm{A}$ is indeed obeyed; see \Eq{wz-scaling}. This suggests that Runs~C and G (green symbols in \Fig{presults_Cm_long_1panel}) with $k_0=10$ and 2, respectively, and with $\mathrm{Ma}_\mathrm{M0}=0$ and $\mathrm{Ma}_\mathrm{M1}=0.005$ also experience magnetically assisted vorticity production.

In \Fig{ABC_large}, we present visualizations of $\nab\cdot\uu$, $\BB^2$, and $\ww^2$ on the periphery of the computational domain for Run N at three different times. We see that the structures reflect the presence of shocks extending over major parts of the domain. This is especially clear in the plots of the local vorticity density.

\section{Conclusions} \label{sec:Conclusions}

Acoustic turbulence is common throughout astrophysics, arising naturally from gravitational accelerations or barotropic pressure fluctuations. In this work, we have used numerical simulations to study the production of vorticity in isothermal, decaying acoustic turbulence, focusing on the role of magnetic fields.

We find that without magnetic fields, acoustic turbulence obeys a Kolmogorov-type phenomenology, with a nondimensional Kolmogorov prefactor of $C_K \approx 6.$ This is significantly larger than the standard Kolmogorov constant for vortical turbulence, which is around 1.6. The presence of a magnetic field lowers this value to around 2--3 for most of our runs, although the universality of this prefactor remains uncertain, as we occasionally observe larger values. 

Magnetic fields also influence the partitioning between the acoustic and vortical components of the turbulence.  When a non-force-free magnetic field is added, the Lorentz force produces vorticity with a kinetic energy spectrum that is close to equipartition with the magnetic energy spectrum in the upper part of the inertial range. The turbulence also begins to resemble vortical turbulence, developing a spectrum that is nearly in equipartition with the magnetic energy spectrum at high wavenumbers.  Our simulations reproduce this process, in agreement with earlier findings \citep{Kahn+12}.

We also show that, even if the magnetic field is force free, it is still able to produce vorticity by converting acoustic energy into vortical kinetic energy. This conversion is most efficient when the acoustic component has significant contributions from large length scales and when the field is strong. The amplitude of the vortical component in this case is expected to scale quadratically with the magnetic field and linearly with the strength of the initial acoustic component. This scaling is confirmed by our simulations, particularly in Runs~N and P, where a strong imposed magnetic field ($\mathrm{Ma}_\mathrm{M0}=1$) converts acoustic energy into vortical energy. Even in the case of a turbulent magnetic field, the same scaling holds, though the required field strength is much weaker ($\mathrm{Ma}_\mathrm{M1}=0.005$). 

The implications of our findings extend to cosmology, particularly to the early Universe. During the radiation-dominated era, the gas obeys an ultrarelativistic equation of state, where the pressure is proportional to the density, similar to isothermal flows. The sudden generation of acoustic turbulence, for example from cosmological phase transitions \citep{Turner+92, Hindmarsh+15}, could be converted into vortical turbulence by a magnetic field. Such a field might have been produced either during inflation or during the subsequent reheating era just prior to the radiation-dominated era. This could play a significant role in shaping the dynamics in the early Universe, particularly the generation of vortical turbulence from initially acoustic fluctuations.

\begin{acknowledgments}
We thank Liubin Pan for helpful discussions that improved the manuscript.
We acknowledge inspiring discussions with the participants of
the program on ``Turbulence in Astrophysical Environments'' at the Kavli
Institute for Theoretical Physics in Santa Barbara.
This research was supported in part by the
Swedish Research Council (Vetenskapsr{\aa}det) under Grant No.\ 2019-04234,
the National Science Foundation under Grants No.\ NSF PHY-2309135,
AST-2307698, and NASA Awards 80NSSC22K0825 and 80NSSC22K1265. 
We acknowledge the allocation of computing resources provided by the
Swedish National Allocations Committee at the Center for
Parallel Computers at the Royal Institute of Technology in Stockholm.

\vspace{2mm}\noindent
{\em Software and Data Availability.}
The source code used for the simulations of this study,
the {\sc Pencil Code} \citep{PC}, is freely available on
\url{https://github.com/pencil-code/Acoustic}.
The simulation setups and corresponding input
and reduced output data are freely available on
\dataset[http://doi.org/10.5281/zenodo.14968754]{http://doi.org/10.5281/zenodo.14968754}
\end{acknowledgments}

\bibliography{ref}{}
\bibliographystyle{aasjournal}

\appendix
\label{Spectral-comparisons}

\begin{figure}[b!]
\includegraphics[width=3.2in]{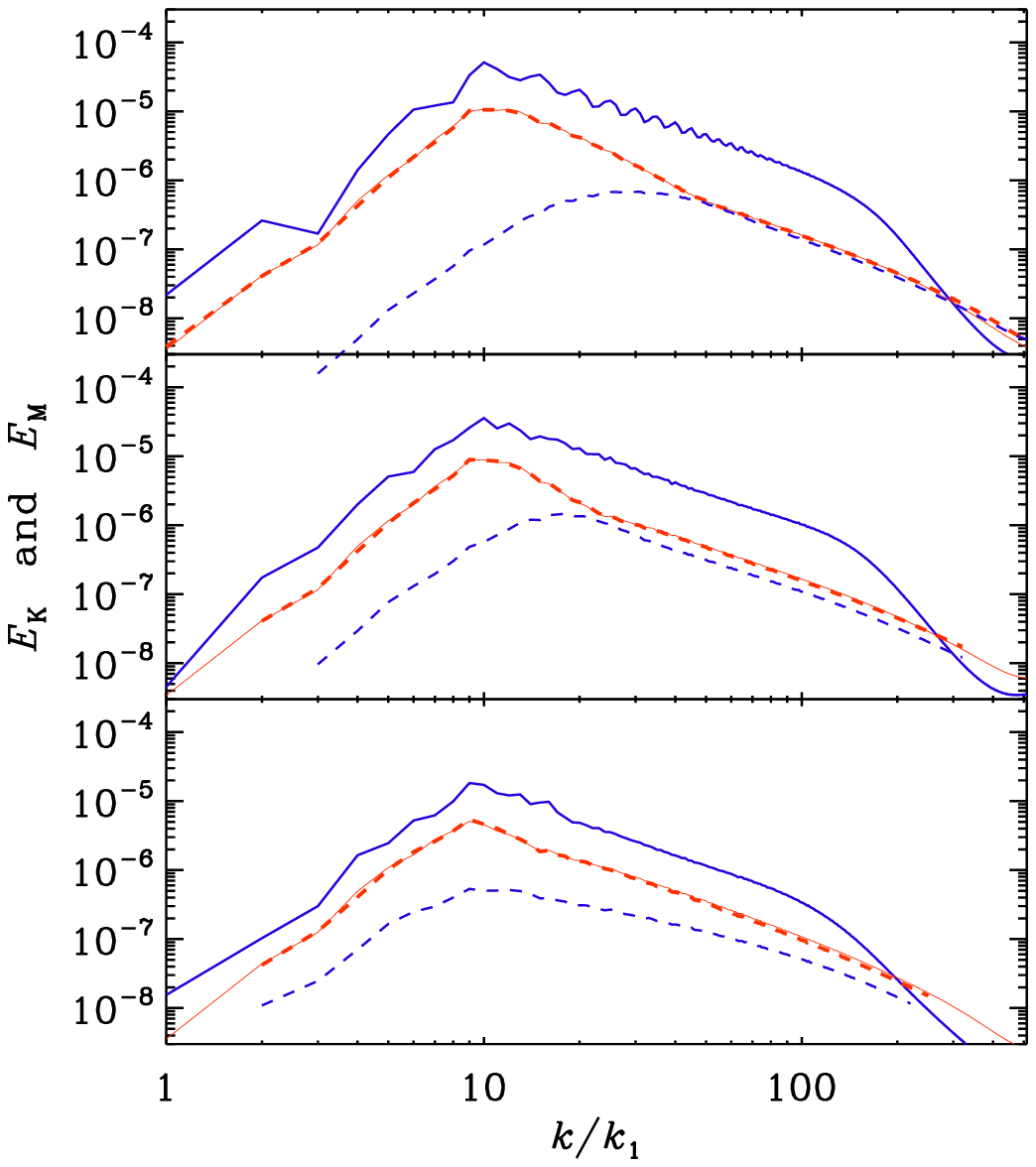}
\vspace{-0.05in}
\caption{
Comparison of kinetic (blue lines) and magnetic (red lines) energy spectra for Runs~C (solid lines) and E (dashed lines), runs with and without initial turbulence, at times 2.5, 7.5, and 25.
\label{ppower_comp}}
\vspace{-0.1in}
\end{figure}

\begin{figure}[h!]
\includegraphics[width=3.2in]{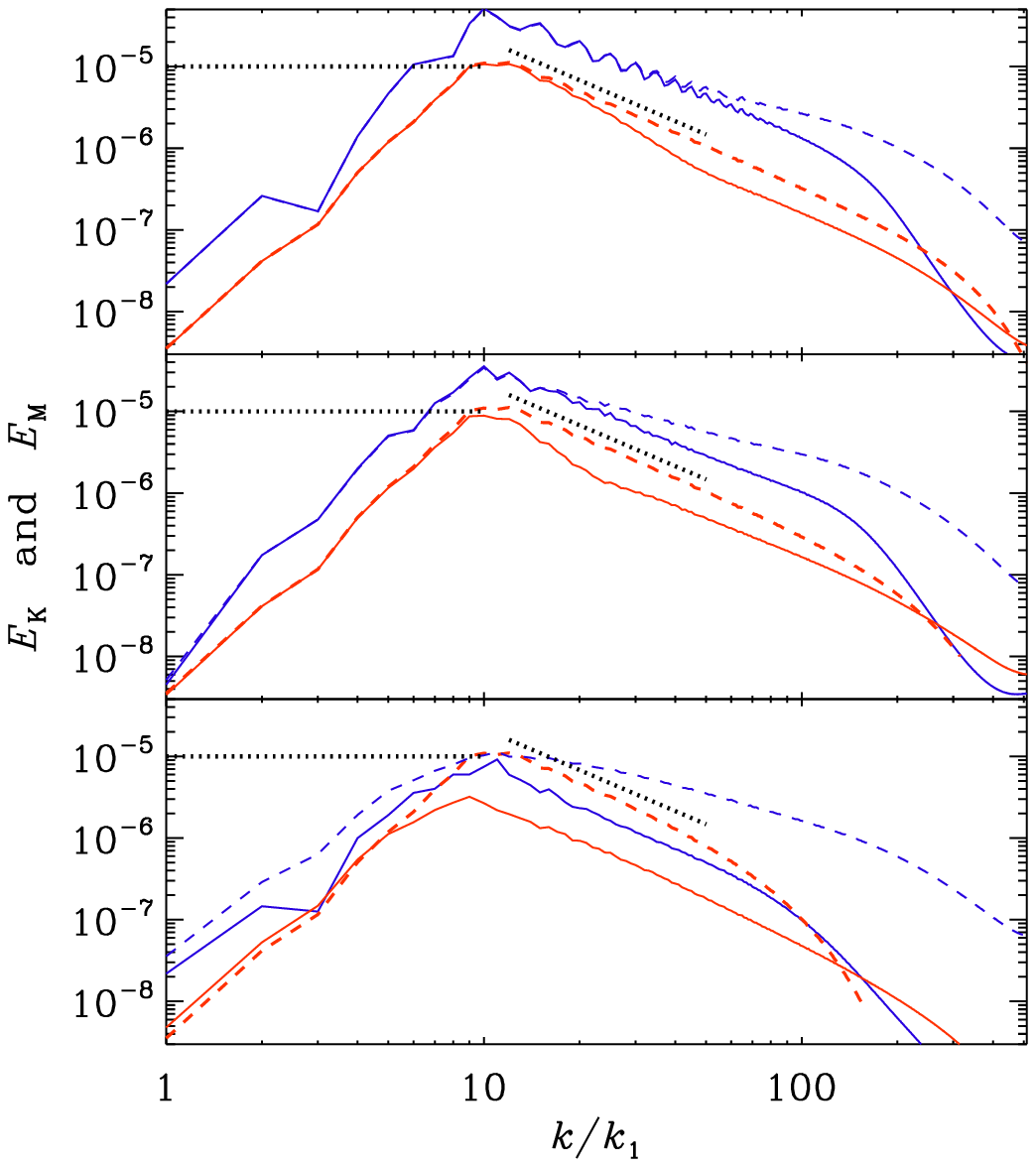}
\vspace{-0.05in}
\caption{
Similar to \Fig{ppower_comp}, but for Runs~C (solid lines)
and D (dashed lines, $\iota=0$, i.e., no induction) at times 2.5, 7.5, and 60.
The black dotted lines provide fixed reference values in each panel.
\label{ppower_comp_mag}}
\vspace{-0.2in}
\end{figure}

\begin{figure}[h!]
\includegraphics[width=3.2in]{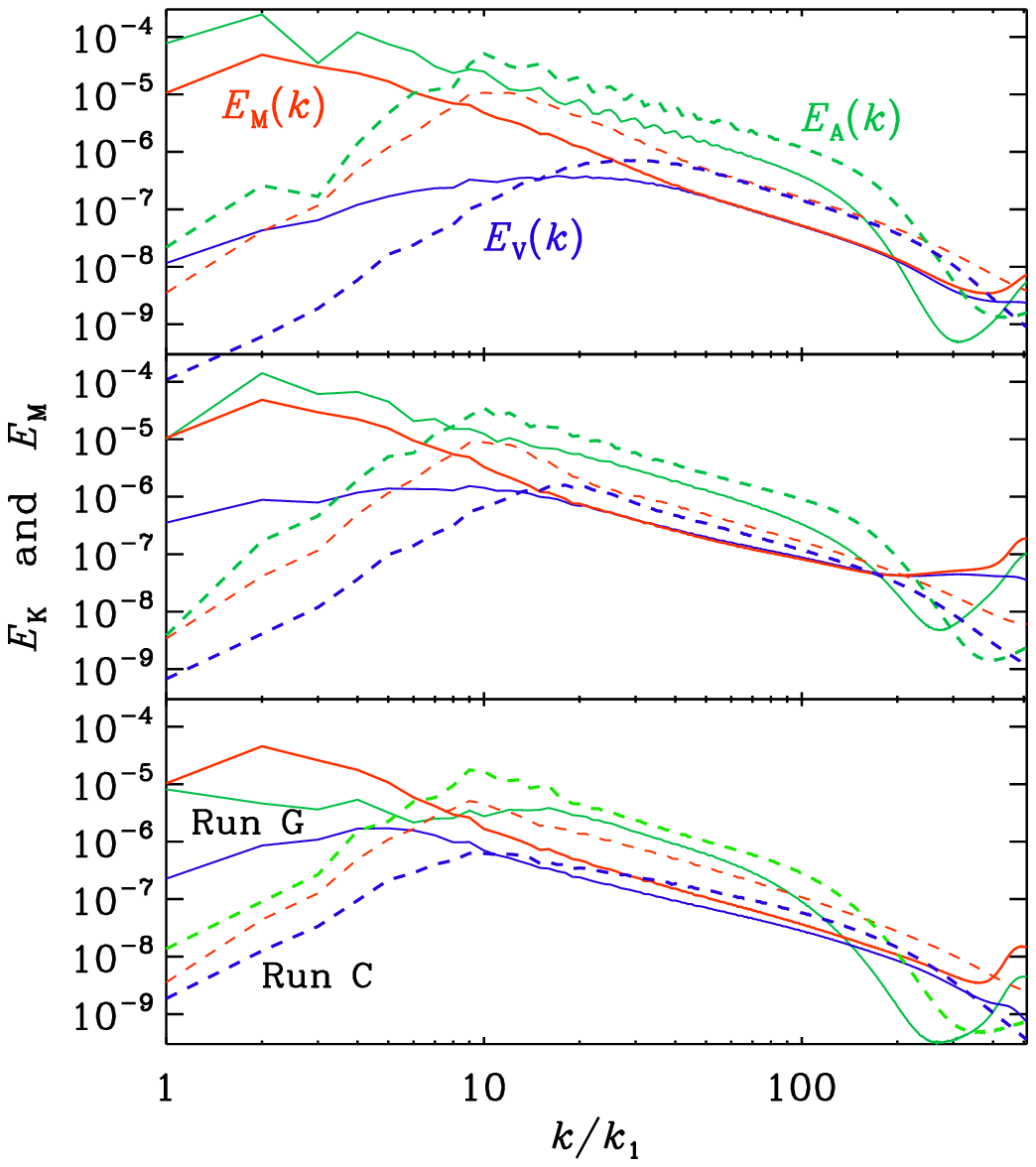}
\vspace{-0.05in}
\caption{
Comparison of acoustic (green lines), vortical kinetic (blue lines), and magnetic (red lines)
energy spectra for Runs~C (dashed lines) and G (solid lines) at times 2.5, 7.5, and 25.
\label{ppower_comp_kf2}}
\vspace{-0.2in}
\end{figure}

At the end of \Sec{sec:CK}, we mentioned spectral comparisons of Run~C with three other runs (Runs~E, D, and G).
In \Fig{ppower_comp}, we compare the kinetic and magnetic energy spectra for Runs~C and E, i.e., with and without initial turbulence. We see that turbulence is gradually generated by the magnetic field, but there is hardly any effect on the magnetic energy spectra.

In \Fig{ppower_comp_mag}, we show the resulting spectra for a case
where the induction term, $\uu\times\BB$, has been suppressed in
\Eq{dAdt}, i.e., $\iota=0$, and we just solve the diffusion equation, $\partial\AAA/\partial t=\eta\nabla^2\AAA$.
The magnetic field then decays preferentially at high wavenumbers, where magnetic diffusion is the strongest. This is evident from a premature cutoff of the magnetic energy spectrum.
The vortical part of the kinetic energy spectrum now seems to show a very strong bottleneck, but the acoustic part does have a plateau at a low level and a small bottleneck.
This suggests that the initial energy in the acoustic component is
unimportant for the dynamics of the magnetic field.
Moreover, the vorticity production by the magnetic field is largely
independent of the initial energy in the irrotational component.

In \Fig{ppower_comp_kf2}, we compare magnetic and kinetic energy spectra
for the vortical and acoustic components for Runs~C with G at three different times.
We see that, at later times, Run~G suffers a loss of kinetic energy
in the acoustic components along with a gain of kinetic energy in the
vortical component.
This energy exchange occurs around the wavenumber $k/k_1=2$.

\end{document}